\documentstyle[12pt]{article}
\input psfig

\def\sec{\,{\rm sec}}

\def\rcm{\,{\rm cm}}

\def\Mpc{\,{\rm Mpc}}

\def\cmm2{{\,\rm cm^{-2}}}
\def\cm2{{\,{\rm cm}^2}}
\def\cmm3{{\,{\rm cm}^{-3}}}
\def\gcmm3{{\,{\rm g\,cm^{-3}}}}
\def\kms{\,{\rm km\,s^{-1}}}

\def\fun#1#2{\lower3.6pt\vbox{\baselineskip0pt\lineskip.9pt
  \ialign{$\mathsurround=0pt#1\hfil##\hfil$\crcr#2\crcr\sim\crcr}}}

\begin{document}
\pagestyle{empty}
\begin{center}
\bigskip

\rightline{astro-ph/9811364}
\rightline{presented at {\it Great Debate:  Cosmology Solved?}}
\rightline{October 4, 1998, Baird Auditorium}
\rightline{Smithsonian Natural History Museum, Washington, DC}
\rightline{to appear in {\it Proc. Astron. Soc. Pacific}, February 1999}

\vspace{1in}
{\Large \bf COSMOLOGY SOLVED?  QUITE POSSIBLY!}
\bigskip

\vspace{.2in}
Michael S. Turner\\

\vspace{.2in}
{\it Departments of Astronomy \& Astrophysics and of Physics\\
Enrico Fermi Institute, The University of Chicago, Chicago, IL~~60637-1433}\\

\vspace{0.1in}
{\it NASA/Fermilab Astrophysics Center\\
Fermi National Accelerator Laboratory, Batavia, IL~~60510-0500}\\

\end{center}

\vspace{.3in}
\centerline{\bf ABSTRACT}
\bigskip

The discovery of the cosmic microwave background (CMB) in 1964 by Penzias
and Wilson led to the establishment of the hot
big-bang cosmological model some ten years later.
Discoveries made in 1998 may ultimately have as profound an
effect on our understanding of the origin and evolution
of the Universe.  Taken at face value, they confirm the basic
tenets of Inflation + Cold Dark Matter, a bold and expansive theory
that addresses all the fundamental questions left
unanswered by the hot big-bang model and holds that
the Universe is flat, slowly moving elementary particles provide the
cosmic infrastructure, and quantum fluctuations seeded all
the structure seen in the Universe today.
Just as it took a decade to establish the hot big-bang model
after the discovery of the CMB, it will likely take another
ten years to establish the latest addition to the standard
cosmology and make the answer to ``Cosmology Solved?'', ``YES!''
Whether or not 1998 proves to be a cosmic milestone,
the coming avalanche of high-quality cosmological data
promises to make the next twenty years an extremely exciting
period for cosmology.

\pagestyle{plain}
\setcounter{page}{1}
\newpage

\centerline{DEDICATION}
\begin{quote}

This article is dedicated to the memory of a great scientist,
my cosmological mentor, and my very dear friend, David N. Schramm.
David had been scheduled to face Jim Peebles in this Great Debate;
after his tragic death in a plane crash last December,
I agreed to take his place in this event which is now dedicated to
his memory.  Throughout his career David was ``bullish''
on cosmology and for a number of years he had been speaking about
the coming Golden Age in cosmology, where a flood of cosmological data
would test the bold ideas that blossomed from the connection
between the Inner Space of elementary particles and the Outer
Space of cosmology which he helped to pioneer.  I am certain that
he would have enjoyed his role in this celebration of cosmology,
and my hunch is that he would have answered the question,
``Cosmology Solved?'', with the answer that I have.

\end{quote}
\vskip 1in

\section{When is Cosmology Solved?}

Cosmology is the scientific study of the origin and evolution of
the Universe, and the word itself derives from the Greek, cosmos,
meaning order.  From my perspective as a particle cosmologist,
I would say that Cosmology
is Solved when we explain and understand the basic features of the Universe,
those which define its fundamental character, in terms of
a theory rooted in fundamental physics.

Solving cosmology does not mean the end of
the study of the Universe, nor even the beginning of a less
exciting period of astrophysical inquiry.  An analogy
may be helpful; we have known the laws of quantum mechanics for
more than sixty years, and quantum physics continues to be a vibrant
field of study, as evidenced by recent advances including
the first Bose -- Einstein condensates of atoms,
high-temperature superconductivity, fractional quantum Hall effect,
quantum computing, and quantum interference devices.

The Universe is the most amazing and wondrous ``zoo''
one can imagine, full of all
kinds of interesting objects and a diversity of
phenomena.  Astrophysics is the
scientific pursuit of an understanding of these objects and
phenomena in terms of the laws of physics.  It is
difficult to imagine astrophysics ever being solved.  A list
of today's puzzles is challenging enough to
occupy astrophysicists for decades:
What are the objects that make gamma-ray bursts
and how do they work?;  How do galaxies form stars and light up
the sky?;  How are stars born?; When were the first stars born?;
What mechanism makes stars explode as supernovae?; Most of the
ordinary matter is not in the form of stars, but is dark -- what is it?;
How do planets form?; Is there life elsewhere in the Cosmos?;
How do massive black holes form? What is the origin of the
highest energy cosmic rays? and on and on.  As we
are flooded with data from new ground-based and
space-based observatories and experiments in the coming years and
some of these questions are answered, the list will grow longer,
with new, more interesting questions being added.
Cosmology solved or not,
I am confident that there will be plenty of challenges
for next century's astrophysicists.

The revolution in cosmology triggered by the discovery of the
CMB in 1964 led to the establishment of the
hot big-bang cosmological model as the standard cosmological model
(see e.g., Silk, 1980; or Peebles et al 1991).  I believe
the hot big-bang theory will be viewed as one of the great intellectual
triumphs of the 20th century.  Based upon a simple mathematical
model, the Friedmann -- Lemaitre -- Robertson -- Walker (FLRW)
solution of Einstein's equations, it describes accurately the evolution
of the Universe from a fraction of a second after the bang until today.  As
discussed by Silk, the FLRW model stands upon three experimental pillars:
the observed expansion of the Universe; the existence of
the cosmic microwave background (CMB) radiation; and the abundance
pattern of the light elements D, $^3$He, $^4$He, and $^7$Li
produced seconds after the bang in a sequence of nuclear reactions
known as big-bang nucleosynthesis (BBN).

As successful as the FLRW cosmology is, there are a number of fundamental
questions that it leaves unexplained.  Here is the list of questions that I
believe must be addressed before we can say ``Cosmology Is Solved'':

\begin{itemize}

\item Origin of the expansion and definitive measure of the present
expansion rate $H_0$ (Hubble's constant).

\item Origin of the heat in the Universe and a precise
measure of the present temperature of the CMB.

\item  Full accounting of matter and energy in the Universe.  From
such an accounting one can infer the present rate of deceleration
(or acceleration) of the expansion and the geometry of the Universe.

\item Understanding of the origin of the density inhomogeneities that
seeded all the structure seen in the Universe today.

\item Understanding of the origin of ordinary matter and particle
dark matter.

\item Understanding of the dynamite behind the big bang.  The
term ``big-bang theory'' is a misnomer -- it is not a theory of the
big-bang event, but rather, of the events thereafter.

\item Understanding of the regularity of the Universe, as
evidenced by the uniformity of the CMB (temperature variations
across the sky of less than one part and in $10^4$ and the
statistically homogeneous distribution of galaxies).

\item Description of the history of the Universe from the big-bang
event on.

\end{itemize}

As I will discuss in detail in Section 3, Inflation + Cold Dark Matter
is a theory that addresses all of these questions as well as extending
our understanding of the Universe back to times as early as
$10^{-32}\sec$.  Its fundamental predictions are that the Universe
is spatially flat, that the bulk of the matter exists in the
form of slowly moving elementary particles (and not the stuff
that we are made of), and that diversity of structure we see
in the Universe today, from galaxies to the great walls of
galaxies (Geller and Huchra, 1989), arose from quantum mechanical
fluctuations on subatomic scales.

\section{1998 -- A Most Memorable Year for Cosmology}

1998 saw the first plausible, complete accounting of matter
and energy in the Universe; a precision determination of
the density of ordinary matter; and other strong evidence supporting
the basic tenets of Inflation + Cold Dark Matter (CDM).  If subsequent
data and observations confirm and strengthen the case, I believe
we will ultimately refer to 1998 as a turning point in cosmology
as important as 1964.

\begin{figure}
%\vspace{3in}
\centerline{\psfig{figure=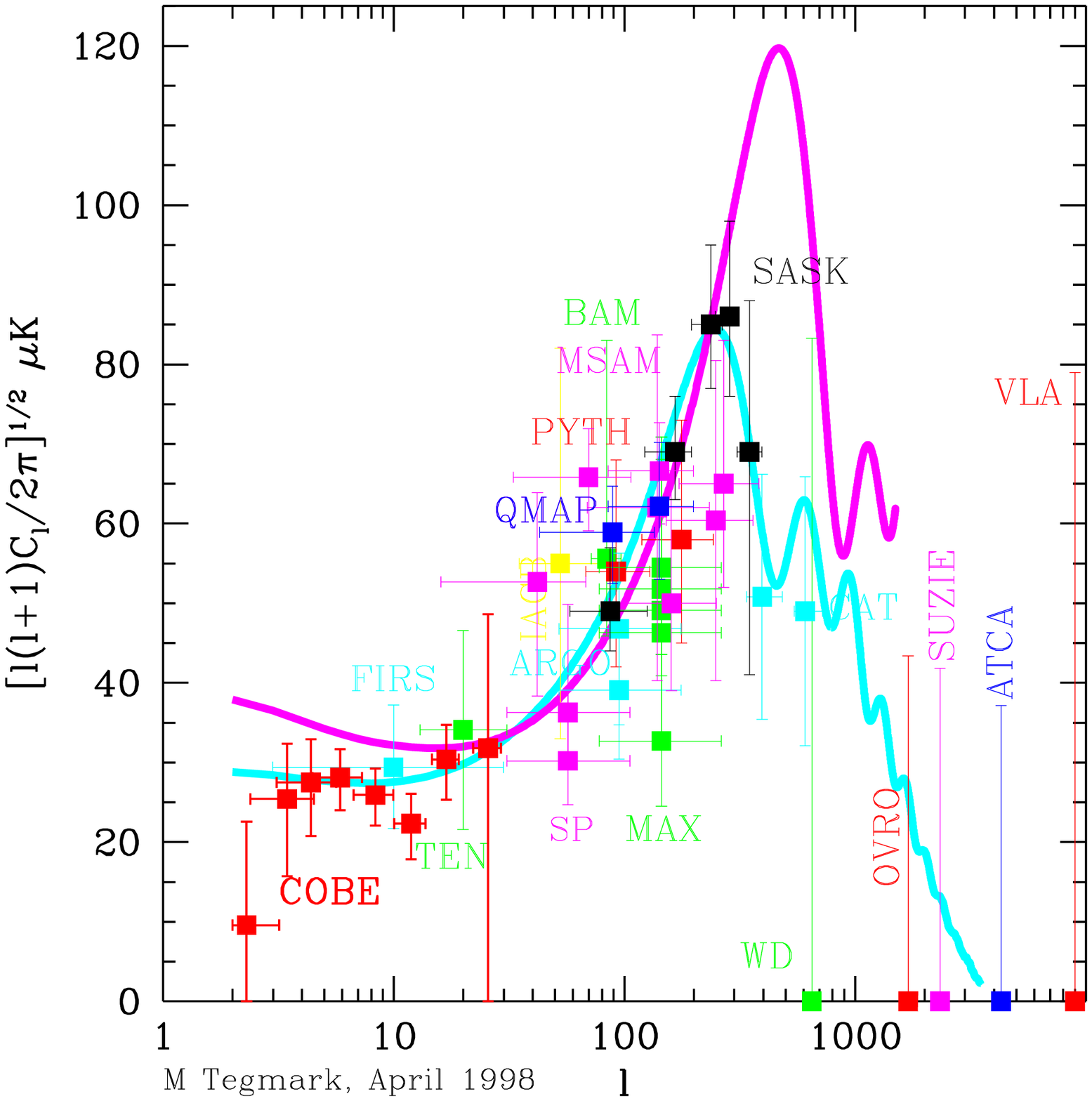,width=4in}}
\caption{Summary of all CMB anisotropy measurements, where
the CMB temperature variation across the sky has been expanded
in spherical harmonics, $\delta T(\theta , \phi ) = \sum_i a_{lm}Y_{lm}$
and $C_l \equiv \langle |a_{lm}|^2\rangle$.  In simple
language, this plot shows the size of the temperature variations
between two points on the sky separated by angle $\theta$
(ordinate) vs. multipole number $l=200^\circ / \theta$
($l=2$ corresponds to $100^\circ$, $l=200$ corresponds to $\theta = 1^\circ$,
and so on).  The curves illustrate the predictions of CDM models
with $\Omega_0 = 1$ (curve with lower peak) and $\Omega_0 =0.3$ (darker
curve).  Note:  the preference of the data for a flat Universe, and
the evidence for the first of a series of ``acoustic peaks.''
The presence of these acoustic peaks is a key signature
of the density perturbations of quantum origin predicted by inflation
(Figure courtesy of M. Tegmark).}
\label{fig:cbr_today}
\end{figure}

As discussed by Silk, the small variations in the CMB temperature
across the sky have the potential to determine the geometry of
the Universe and thereby $\Omega_0$, the fraction of critical
density contributed by all forms of matter and energy.  [Note,
the curvature radius of the Universe and $\Omega_0$ are related
by, $R_{\rm curv}^2 = H_0^{-2}/|\Omega_0 - 1|$, so that $\Omega_0 =1$
corresponds to a flat geometry, $\Omega_0<1$ corresponds to
negatively curved (open) geometry, and $\Omega_0 > 1$ corresponds
to positively curved (closed) geometry.]  In a flat Universe
the differences in temperature between points on the sky are greatest when the
two points are separated by about one degree; in an open
Universe, they are greatest when the separation is smaller than one degree.
The data that now exist indicate that the differences are
indeed greatest on the one-degree scale (see Fig.~\ref{fig:cbr_today}).
More measurements
are being made -- from balloons, in Antarctica, and on the
Atacama Plateau in Chile -- with definitive measurements to come from
two new satellites: NASA's MAP (launch 2000) and ESA's Planck
Surveyor (launch (2007).  If the trend in the results continues,
then we have determined that the Universe is
spatially flat and $\Omega_0=1$.

These same measurements of the temperature variations of the
CMB temperature across the sky also provide important information
about the matter inhomogeneities that
seeded all the structure we see in the Universe today.
Recall, that as Silk explained, the pattern of hot and cold spots on
the microwave sky arises due to the lumpy distribution of matter, and
in this way the CMB is a snapshot of the Universe at 300,000 years
after the beginning (CMB photons come directly to us from
their last scattering at this time).  The first mapping of
the distribution of matter in
early Universe using the CMB was done by
NASA's COBE satellite (Smoot et al, 1992); since then,
many other experiments, with better angular resolution, have ``enhanced
the picture.''  At present, the pattern of hot and cold spots
indicate, though do not yet prove, that the primeval lumpiness
had the following characteristics:  Gaussian fluctuations in the
curvature of the Universe with an approximately scale-invariant
spectrum (see Fig.~\ref{fig:cbr_today}).
This is precisely what inflation predicts.

\begin{figure}
%\vspace{3in}
\centerline{\psfig{figure=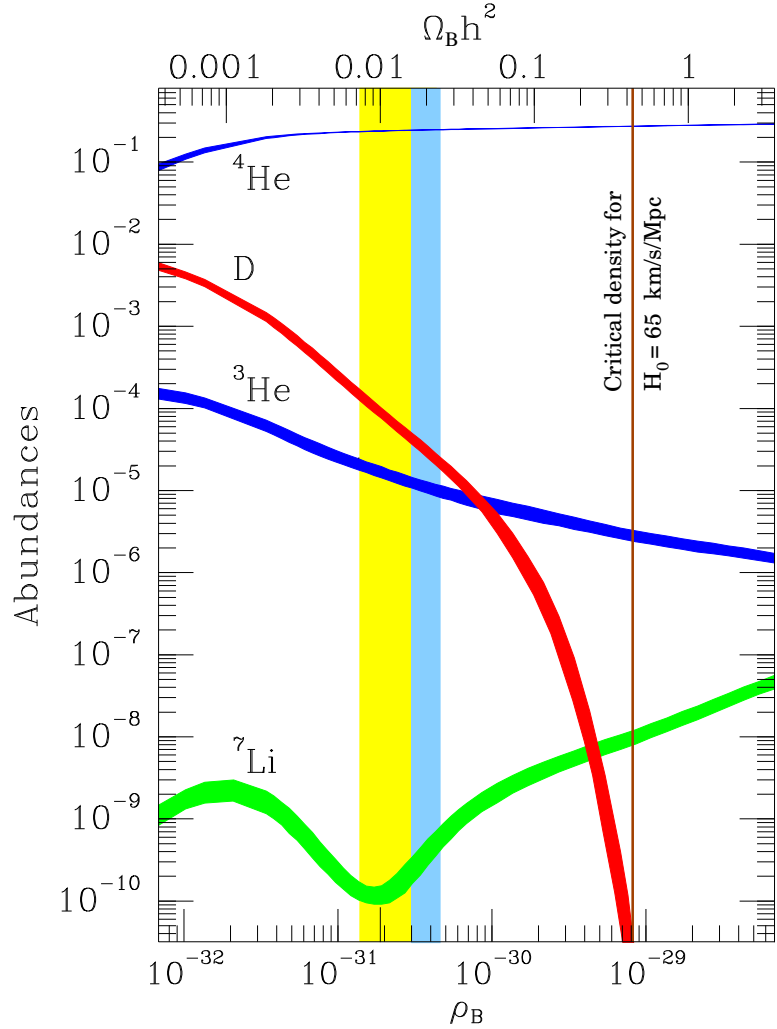,width=4in}}
\caption{Predicted abundances of $^4$He, D, $^3$He, and $^7$Li
(relative to hydrogen) as a function of the density of ordinary
matter (baryons).
The full band denotes the concordance interval based upon all
four light elements that dates back to 1995.  The darker portion highlights
the determination of the density of ordinary matter based upon
the recent measurement of the primordial abundance of
deuterium (Burles \& Tytler, 1998a,b), which implies
that ordinary matter contributes 5\% of the critical density.
}
\label{fig:bbn}
\end{figure}

Of the four light elements made in the big bang, the yield of
deuterium is most sensitive to the density of ordinary matter
(baryons), and for that reason David Schramm and
I nicknamed it ``the baryometer'' (Schramm \& Turner, 1998).
Deuterium is also ``very fragile,'' easily destroyed
by nuclear reactions in stars.  The abundance of deuterium
in our local neighborhood has been known for more than twenty
years, but because about half of the material around us has been
through stars, we cannot interpret the locally measured
abundance as the primordial, big-bang
abundance.  This year, David Tytler and
Scott Burles (Burles \& Tytler, 1998a,b) measured
the deuterium abundance in very distant hydrogen clouds.
Because these clouds are so distant, we are seeing them
at an early time, before stars have destroyed their
deuterium.  Their measurement of the primeval
deuterium abundance pegs the contribution
of ordinary matter to be 5\% of the critical density (for a
Hubble constant of $65\kms\Mpc^{-1}$), with a precision of
better than $\pm 0.5\%$ (see Fig.~\ref{fig:bbn}).

In addition to pinning down the amount of ordinary matter, the Tytler -- Burles
measurement has allowed us to determine the total amount of matter,
using a very clever technique pioneered by Simon White and his
colleagues (White et al, 1993).  They argue that clusters of galaxies, by virtue of
their large size (tens of millions of light years in size and thousands
of galaxies in number), represent a fair sample of matter in the Universe.
Thus, the ratio of baryons to matter in clusters, together with
the big-bang determination of the average density of baryons in
the Universe serve to pin down the average matter density in
the Universe:
\begin{equation}
\Omega_{\rm Matter} = {\Omega_{\rm Baryons}\over {\rm ratio\ of\ baryons\
to \ matter\ in\ clusters}} = {5\%\over 13\%} = 40\%\pm 10\%
\end{equation}
Most of the baryons in clusters exist in the form of hot, x-ray
emitting gas and the intensity of x-ray emission allows a determination
of the amount of ordinary matter.  The total cluster mass can be
determined in a number of ways:  by direct mapping of the
mass by gravitational lensing, by measuring the x-ray temperature,
and by measuring the random
motion of galaxies; all three give the same answer.
The ratio of baryons to matter has been determined
to be $13\%\pm 1.5\%$.  This implies that matter contributes $40\%\pm 10\%$
of the critical density.  (Other determinations of the matter density
indicate a similar value.)

The determination that matter contributes 40\% of the
critical density has two implications.  First,
as Silk discussed, most of the
matter in the Universe is not the stuff we are made out of, which
follows from the inequality, $\Omega_{\rm Matter} > \Omega_{\rm
Baryon}$.  That is, the matter that provides the cosmic infrastructure
and holds the Universe together is something exotic (or, perhaps
we should say that we are made of something more exotic).  That
is a remarkable fact.
As I will discuss, the most promising idea is that this exotic
matter is relic elementary particles
left over from the earliest moments of creation.

The second, equally profound implication follows from
the inequality, $\Omega_0 > \Omega_{\rm Matter}$.  This implies
the existence of a form of matter or energy that does not clump
(as evidenced by the fact that it is not found in clusters) and
yet contributes 60\% of the critical density.
The fact that this funny component to the energy density does
not clump and contributes so much of the critical density implies
that it must be very elastic (the technical term is negative pressure)
and leads to a striking prediction:  that the expansion of the
Universe should be speeding up, rather than slowing down!

This requires further explanation:
According to Einstein's theory of general relativity, the source
of gravity is energy density + three times the pressure (Note:
matter and energy are equivalent, related by $E=mc^2$; Newton's
theory of gravity holds that the source of gravity is matter alone.)
If the pressure is sufficiently negative -- and it must be if
the additional component is to remain smooth -- its gravity is
repulsive!  Because there is so much of this ``funny'' energy,
the net effect of gravity on the expansion of the Universe
is repulsive and so the Universe should be speeding up, rather than slowing
down!  Sandage's famous deceleration parameter embodies all
of this, $q_0 = {\Omega_0\over 2}[1+3p/\rho]$ is negative
if $p<-\rho /3$.

(An aside:  the fact that energy density + three times
pressure is the source of gravity in Einstein's theory also leads
to the prediction of black holes.  An object with very strong gravity
needs great pressure to balance gravity; objects with stronger
and stronger gravity --  i.e., stars of greater and greater mass --
need more and more pressure; eventually, the pressure becomes
counterproductive, producing more gravity, resulting in a black hole.)

The prediction of accelerated expansion was confirmed
in 1998 and completed the accounting of matter and energy in the Universe.
By studying the relationship between the distances and
velocities of distant galaxies by using exploding
stars (supernovae of type Ia, or SNeIa) within them
as standard candles, two teams (the Supernova
Cosmology Project led by Saul Perlmutter and the High-redshift
Search Team led by Brian Schmidt) found
evidence that the Universe is speeding up, rather than slowing down
(Perlmutter et al, 1998; Riess et al, 1998).
Let me explain their results by telling you what they expected to
find and what they actually did find.  Because the expansion of the
Universe is simply a scaling up of all distances, if we could measure galaxy
velocities and positions today, they would obey exactly
Hubble's law:  recessional velocity proportional
to distance ($v=H_0 d$).  However,
as we look far out into space, we look back in time (light travels at
a finite speed), and so we are viewing distant galaxies at earlier
and earlier times.  If the Universe is slowing down, distant galaxies
should be moving faster than Hubble's law predicts.  These two
groups found the opposite:  distant galaxies are moving more slowly
than Hubble's law predicts.  The expansion is speeding up.

The simplest interpretation of their results is that vacuum energy,
with pressure equal to minus its energy density, contributes about
60\% of the critical density, making consistent the determinations
$\Omega_0 =1$ and $\Omega_{\rm Matter} = 0.4$.
Vacuum energy is the modern term
we use for Einstein's cosmological constant.  It corresponds to
the energy associated with the very lively quantum vacuum -- pairs
of particles borrowing enough energy to exist for a fleeting instant
and then disappearing again.

\begin{figure}
%\vspace{3.5in}
\centerline{\psfig{figure=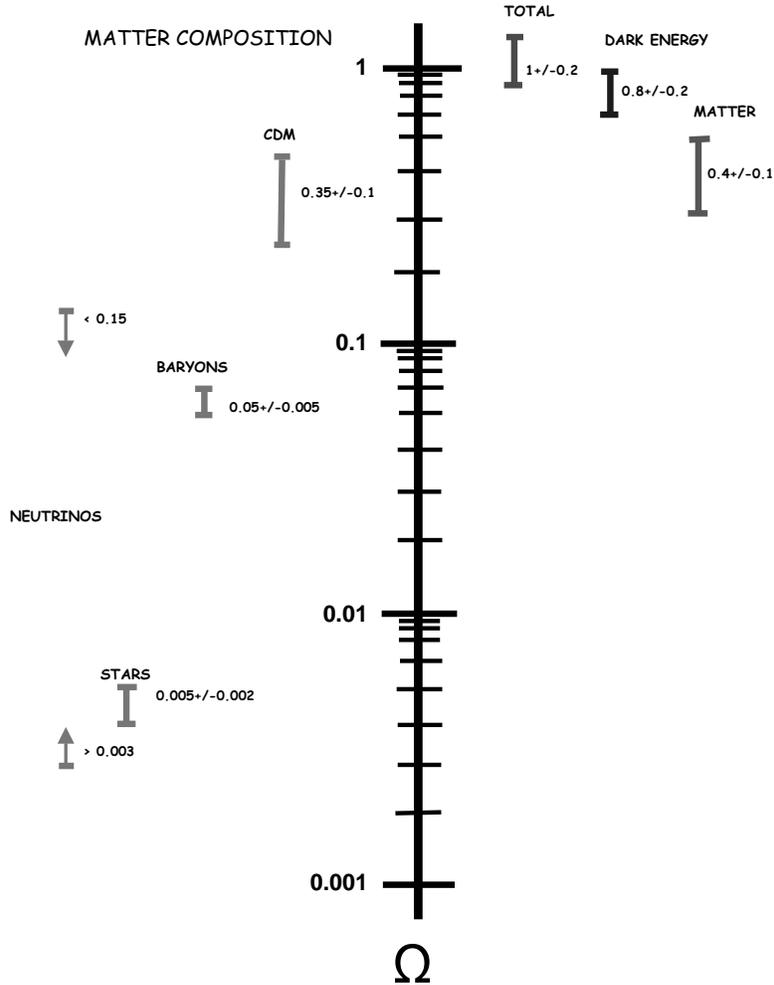,width=4in}}
\caption{Summary of matter/energy in the Universe.
The right side refers to an overall accounting of matter
and energy; the left refers to the composition of the matter
component.  The upper limit to mass density contributed
by neutrinos is based upon the failure of the hot dark
matter model of structure formation
and the lower limit follows from the
evidence for neutrino oscillations (Fukuda et al, 1998).
Here $H_0$ is taken to be $65\kms\Mpc^{-1}$.
}
\label{fig:omega}
\end{figure}

To summarize the accounting of matter and energy in the Universe
(in units of the critical density): neutrinos, 0.3\% or greater;
bright stars, 0.5\%; total amount of ordinary matter
(baryons), 5\%; relic elementary particles, 35\%; and vacuum
energy (or something similar), 60\%; for a total of 100\% of
the critical density (see Fig.~\ref{fig:omega}).
While this accounting is not definitive yet,
measurements that are being made and will be made in the
next years could firm it up.

(Only about one-tenth of the baryons are ``visible'' in the
form of bright stars.  Further, because there are almost as many
neutrinos left over from the big bang as there are CMB photons and
because evidence now exists that at least two of the neutrino species have
mass, albeit very tiny (Fukuda et al, 1998),
we can infer that neutrinos contribute as much mass,
and perhaps even more, than stars do.  However, both neutrinos
and bright stars are small contributors to the total mass
in the Universe.)

\section{Inflation + Cold Dark Matter}

%% three key predictions
%% varieties of CDM

In addition to providing an account of events from a fraction
of a second (the time of big-bang nucleosynthesis) to the present,
the hot big-bang cosmology, supplemented by the standard model of
particle physics and other advances in our understanding
of the fundamental particles and their interactions,
provides a firm foundation for speculations
about much earlier times, back to $10^{-43}\sec$
after the beginning (earlier, the quantum nature
of gravity and possibly space-time itself must be considered).

These speculations involve a crucial connection between the inner
space of elementary particles and the outer space of cosmology.
That connection is simple:   when the Universe
was young, it was a hot soup of the fundamental particles of
nature, quarks, antiquarks, electrons, positrons, neutrinos, antineutrinos,
photons, gluons, and other particles.
To understand the earliest moments of creation, one has to
understand the fundamental particles and how they interact
with one another.  The highly successful, standard model
of particle physics provides the information needed to
take us back to about $10^{-11}\sec$; ideas about how the
forces and particles are unified (e.g., supersymmetry,
grand unification and superstring theory) are needed to discuss the
Universe at even earlier times.

The duality of inner space /
outer space connection is also worth noting:  The quark soup of
the early Universe can be recreated at
particle accelerators by colliding high-energy particles together;
the early Universe, with its sea of extremely energetic
particles that are constantly colliding, can be used to
study the forces and particles at energies beyond the reach of
terrestrial accelerators (see e.g., Kolb \& Turner, 1990).
Motivated by interesting and sometimes compelling speculations about fundamental physics
and the unification of the forces of nature, the past fifteen
years have seen much discussion of the earliest history of the
Universe.  These cosmological speculations have allowed cosmologists
to address the most fundamental questions they face; conversely, cosmology
has given particle physicists access to a new laboratory with
virtually unlimited energy.

The most compelling idea to arise from the synthesis
of elementary particle physics with cosmology is Inflation
+ Cold Dark Matter (Guth, 1982; Blumenthal et al, 1984).
It is an expansive paradigm, deeply rooted in fundamental
physics, and it has the potential to extend our understanding of the Universe
back to $10^{-32}\sec$ and to address most of the fundamental questions
poised by the hot big-bang model.  Inflation
+ Cold Dark Matter holds that most of the dark matter consists of
slowly moving elementary particles; that the Universe is flat; and
that the density perturbations that seeded all the structure seen
today arose from quantum mechanical fluctuations on scales of $10^{-23}
\rcm$ or smaller.  It took a while for cosmological observers and experimentalists
to take this paradigm seriously enough to try to disprove it;
now, in the 1990s, it is being tested in a serious way -- and
is passing the tests with flying colors.
(David Schramm played a very crucial role in this regard --
he urged the more conservative observers to take these new
ideas seriously, and with equal fervor, he urged particle cosmologists
to make testable predictions.)
My thesis for this debate is that the first evidence supporting
the fundamental tenets of Inflation + Cold Dark Matter have been
presented this year.

The key feature of inflation -- and the one responsible for its name --
is the tremendous burst of expansion:  In $10^{-32}\sec$ the Universe
grew in size by a factor greater than it has since!  I will not
discuss the details of what caused this burst of expansion; it suffices
to say that it is related to vacuum energy.
This tremendous growth in the size
of the Universe means that all we can see today originated from
an extraordinarily tiny bit of the whole Universe.  A tiny bit
of any space appears smooth and flat -- take the earth for example --
and this leads to the first key prediction of inflation:  the Universe
should appear flat and thus the total energy density should
equal the critical density.  Further, it explains
the large-scale regularity seen today.  (The subsequent
expansion of the Universe does not change the curvature or
regularity; to be very precise, inflation does not predict
an exactly flat Universe, and only predicts that a region
much, much larger than the observable Universe is smooth.)

The quantum world of subatomic particles is in constant turmoil,
fluctuating and changing.  Because we do not live in the subatomic
world, we are unaware of these quantum fluctuations.  However, the
extraordinary burst of expansion stretches quantum fluctuations to
astrophysical scales, making them relevant for the Universe.
And, in a well defined way, this quantum turmoil leads to
the primeval lumpiness in the distribution of matter in the Universe.
The quantum-born, inflation-produced fluctuations are of a form
known as Gaussian scale-invariant curvature fluctuations
(Guth \& Pi, 1982; Hawking, 1983; Starobinskii, 1982; Bardeen,
Steinhardt, and Turner, 1983).  Inflation-produced, quantum fluctuations
in space time itself lead to a relic background of gravitational
waves that are an additional ``smoking gun'' signature of inflation.

On to the cold dark matter part.  Inflation predicts a flat Universe;
that is, total energy density equal to the critical energy density.
Inflation does not predict what form or forms the critical density takes;
we must rely upon astrophysical clues and measurements.
Since ordinary matter (baryons) only contributes about 5\%
of the critical density, and there is good evidence that the
total amount of matter is 40\% of the critical density, there must
be another form of matter in addition to baryons.  The leading
possibility is elementary particles left over from the earliest,
fiery moments.  Because of the high temperatures that existed
early on, the full zoo of elementary particles was represented.
Of interest for cosmology, are particles that are long lived or
stable, and interact sufficiently weakly so that they would
not have annihilated by the present.  Generically, they
fall into two classes -- fast moving, or hot dark matter; and
slowly moving, or cold dark matter.  Neutrinos
are the prime example of hot dark matter -- they move
quickly because they are very light.  Axions and neutralinos
are examples of cold dark
matter. Neutralinos move slowly because they are very heavy
(fifty to five hundred times the mass of a proton)
Axions are extremely
light (one millionth of a millionth the mass of an electron), but
were produced in a very, very cold state (Bose -- Einstein condensate).
Both the axion and neutralino are as of yet hypothetical particles:
they are predicted by theories that unify the forces and particles of
Nature, but they are not yet ruled in or ruled out by experiment.

Motivated by since-refuted experimental evidence
that neutrinos have enough mass to account for
the critical density, hot dark matter was
carefully studied in the 1980s and found wanting (White, Frenk,
and Davis, 1983).
With hot dark matter structure in the Universe
forms from the top down:  large things, like superclusters form
first, and then fragment into smaller objects such as galaxies.  This
is because fast moving neutrinos erase lumpiness
on small scales by moving from regions of greater density
into regions of lower density.
Observations now very clearly indicate that galaxies formed
at redshifts $z\sim 2 - 4$, clusters formed at redshifts
 $z\sim 0 - 1$, and superclusters are just
forming today. So neutrinos are out, at least as the major
component of the dark matter.  This leaves cold dark matter.
(While apparently not a major ingredient in the
cosmic mix, neutrinos may play the role of a needed cosmic spice;
as discussed earlier, there is now experimental evidence
that at least two of the neutrino species are massive.)

Cold dark matter particles cannot move far enough to smooth out lumpiness
on small scales, and so structure forms from the bottom up:  galaxies,
followed by clusters of galaxies, and so on.  The bulk of galaxies should
form around redshifts of $2-4$, followed by clusters at
redshifts $0-1$ and superclusters today.  This is just
what the observations of the young Universe made by the
Keck 10-meter telescopes and the Hubble Space Telescope indicate.

\begin{figure}
%\vspace{3in}
\centerline{\psfig{figure=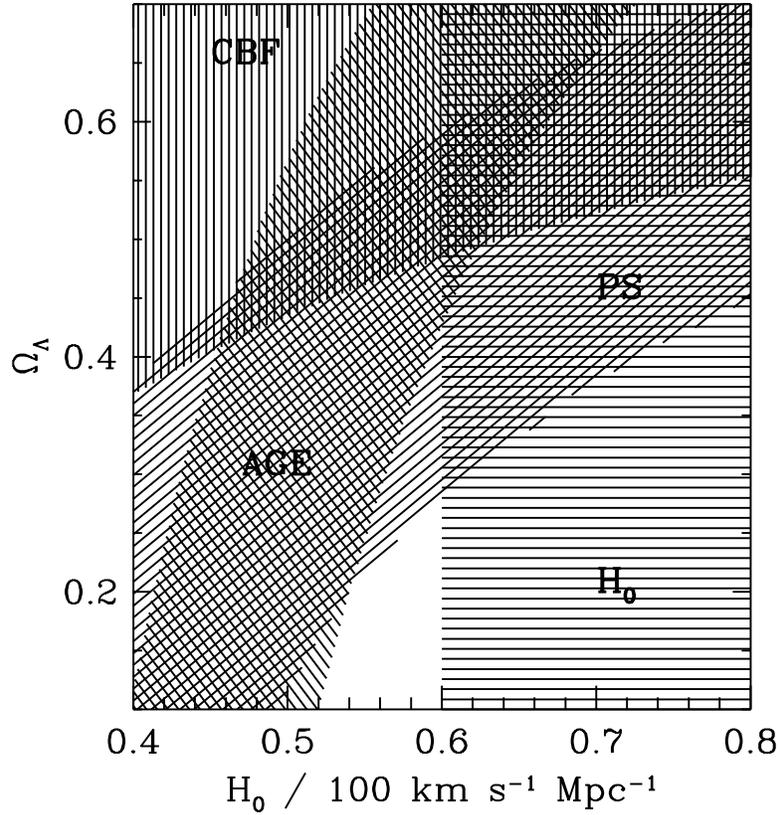,width=4in}}
\caption{Constraints used to determine the best-fit CDM model:
PS = large-scale structure + CBR anisotropy; AGE = age of the
Universe; CBF = cluster-baryon fraction; and $H_0$= Hubble
constant measurements.  The best-fit model, indicated by
the darkest region, has $H_0\simeq 60-65\kms\Mpc^{-1}$ and $\Omega_\Lambda
\simeq 0.55 - 0.65$.  Evidence for its smoking gun signature --
accelerated expansion -- was presented in 1998 by Perlmutter
et al and Riess et al.}
\label{fig:best_fit}
\end{figure}

\begin{figure}
%\vspace{3in}
\centerline{\psfig{figure=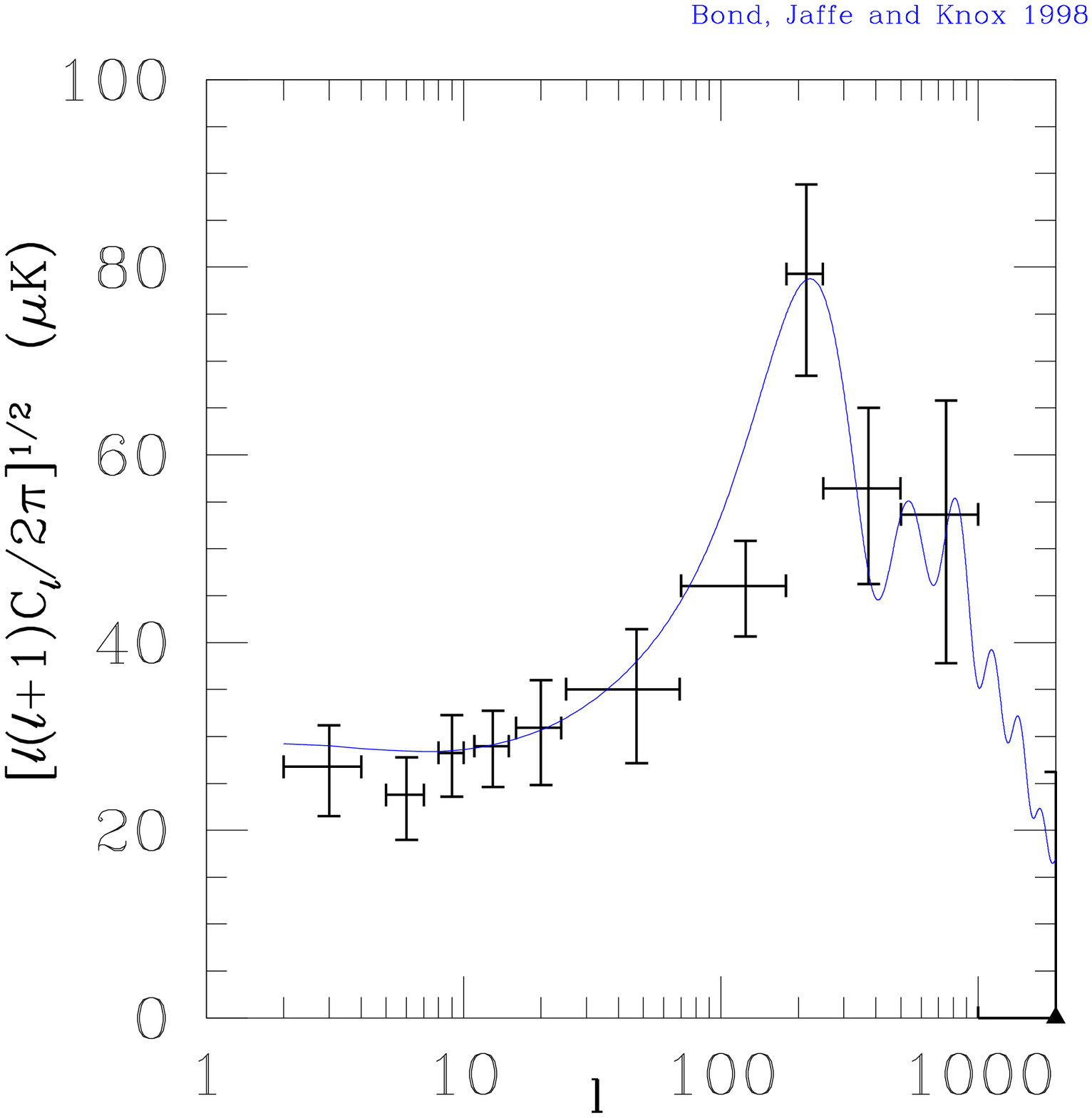,width=4in}}
\caption{The same data in Fig.~1, but
averaged and binned to reduce error bars and
visual confusion.  The theoretical
curve is for the $\Lambda$CDM model with $H_0=65\,{\rm km\,
s^{-1}\,Mpc^{-1}}$ and $\Omega_M =0.4$ (Figure courtesy of L. Knox).
}
\label{fig:cbr_knox}
\end{figure}

The cold dark matter model with a cosmological constant,
referred to as $\Lambda$CDM by the experts,  is consistent with
an enormous body of cosmological and astrophysical data, from
the determinations of the age of the Universe to the pattern
of hot and cold spots in the CMB (see Figs.~\ref{fig:best_fit},
\ref{fig:cbr_knox}, and Turner, 1997; Krauss \& Turner, 1995;
Ostriker \& Steinhardt, 1995).  And now, its dramatic
prediction, that the Universe should be speeding up rather than
slowing down, has been verified (Riess et al, 1998; Perlmutter et al,
1998).  In $\Lambda$CDM the dark energy exists in the form of
spatially constant vacuum energy (Einstein's cosmological constant).
It accounts for 60\% of the
critical energy density, but plays no direct role in the formation of cosmic
structure because it cannot clump.

\section{Cosmology Solved:  The Case for Inflation + Cold Dark Matter}

To make my case that twenty years from now cosmologists will refer
to 1998 as the year Cosmology was Solved, let me return to
my list of necessary elements from the first Section.
Here are the explanations according to Inflation + Cold Dark Matter.

\begin{itemize}

\item Origin of the expansion and definitive measure of the present
expansion rate $H_0$ (Hubble's constant).  {\em The Universe is
still expanding from the inflationary explosion.  Thanks to the
Hubble Space Telescope's calibration of standard cosmological
candles (especially Type Ia supernovae), and techniques based on
gravitational lensing and the influence of hot gas in clusters
upon the cosmic microwave background radiation, we are zeroing in
on the elusive Hubble constant:  all current data are consistent
with $65\pm 5\kms\Mpc^{-1}$ (Madore et al, 1998).}

\item Origin of the heat in the Universe and a precise
measure of the present temperature of the CMB.  {\em The vacuum energy
that drove inflation ultimately decays into radiation (heat), and
according to inflation, the CMB is the primary fossil of inflation!
Thanks to the extraordinary work of John Mather's COBE FIRAS team,
the temperature of the CMB
has been measured to 4 significant figures, as accurately as
the thermometers on the COBE satellite would
permit, $T_0=2.728 \pm 0.002\,$K (Fixsen et al, 1996).}

\item  Full accounting of matter and energy in the Universe.
{\em Inflation predicts that we live in a flat Universe and that
the total energy density is equal to the critical density.
Observations now provide the following accounting:
ordinary matter, 5\%; relic elementary particles, 35\%;
vacuum energy, 60\%, for a total summing to 100\% of the critical
density (see Fig.~\ref{fig:omega}.}

\item Understanding of the origin of the density inhomogeneities that
seeded all the structure seen in the Universe today.  {\em They arose from
quantum fluctuations on subatomic scales that were stretched to
astrophysical size during inflation.  The pattern of hot and cold
spots on the CMB sky are consistent with this prediction
(see Figs.~1 and \ref{fig:cbr_knox}.}

\item Understanding of the origin of ordinary matter and particle
dark matter.  {\em The origin of ordinary matter in the Universe
traces to a slight excess -- part in $10^{10}$ of matter over
antimatter in the early Universe.  As the Universe cooled,
all the antimatter annihilated
with matter, leaving a tiny bit of matter.  Because of the tremendous
heat release at the end of inflation, this tiny excess of matter
over antimatter must
arise after inflation, by interactions among the sea of elementary
particles present.  A framework for understanding this -- called
baryogenesis -- exists and only the details need to be worked out.
The cold dark matter particles remain from the early
moments of creation because they are stable and they are ineffective in
annihilating with one another.}

\item Understanding of the dynamite behind the big bang.
{\em The explosive expansion caused
by vacuum energy (or something similar) is the dynamite behind the
big bang.  Further, in the context of inflationary cosmology, what
we previously called ``The Big Bang,'' which was supposed to be the
creation of the entire Universe, is demoted to ``our big bang''
and the creation of the large, smooth region of the Universe in
which we live.  According to Andrei Linde,
if inflation occurred one, it occurred an infinite
number of times and our bang is but
one of an infinite number (Linde, 1990).}

\item Understanding of the regularity of the Universe.
{\em The portion of the Universe
that we see is very regular because it all originated from an
extraordinarily tiny portion of the Universe.}

\item Description of the history of the Universe from the big-bang
event on.  {\em The events after the vacuum energy of inflation is released
as heat are as in the standard hot big-bang model.}

\end{itemize}

\section{Checklist for the Next Decade}

As I have been careful to stress (far too carefully for a real
debate), the basic tenets of Inflation
+ Cold Dark Matter have not yet been confirmed definitively.
However, a flood of high-quality cosmological data is
coming, and could make the case soon.   Regardless, the flood
of information will make cosmology exciting for the next decade and
beyond.  Here is my version of how ``quite possibly'' becomes ``yes.''

\begin{itemize}

\item Map of the Universe at 300,000 yrs.  COBE mapped
the CMB with an angular resolution of around $10^\circ$;
two new satellite missions, NASA's MAP (launch 2000)
and ESA's Planck Surveyor (launch 2007), will map the
CMB with 100 times better resolution ($0.1^\circ$). From
these maps of the Universe as it existed at a
simpler time, long before the first stars and
galaxies, will come a gold mine of information:
Among other things, a definitive measurement of $\Omega_0$;
a determination of the Hubble constant to a precision of
better than 5\%;
a characterization of the primeval lumpiness; and possible
detection of the relic gravity waves from inflation.
The precision maps of the CMB that will be made are crucial
to establishing Inflation + Cold Dark Matter (see e.g., Bennett
et al, 1997).

\item Map of the Universe today.  Our knowledge of the structure
of the Universe is based upon maps constructed from the positions
of some 30,000 galaxies in our own backyard.  The Sloan Digital
Sky Survey (SDSS, 1998) will
produce a map of a representative portion of the Universe,
based upon the positions of a million galaxies.
The Anglo-Australian Two-degree Field survey will determine the position of
several hundred thousand galaxies (2dF, 1998).  These surveys will
define precisely the large-scale structure that exists today,
answering questions such as, ``What are the largest structures
that exist?''  Together, the CMB map of the young Universe
and the SDSS/2dF map of the Universe today will definitively test the Cold
Dark Matter theory of structure formation, and much more.

\item Cold dark matter.  A key element of theory is the
cold dark matter particles that hold the Universe together;
until we actually
detect cold dark matter particles, it will be difficult to argue
that cosmology is solved.
Experiments designed to detect the dark matter that
holds are own galaxy together are now operating with sufficient
sensitivity to detect both neutralinos and axions (Sadoulet, 1999).
In addition, experiments at particle accelerators (Fermilab
and CERN) will be hunting for the neutralino and its other
supersymmetric cousins.

\item Nature of the dark energy.  If the Universe is indeed accelerating,
then most of the critical density exists in the form of dark energy.  This
component is poorly understood.  Equally puzzling is why it is just
now come to be the dominant component of the mass/energy budget:  its
energy density is constant (or slowly varying) and the matter density
decreases as the volume of the Universe increase, and thus in the
past it was unimportant and in the future matter will be unimportant.
Independent evidence for the existence
of this dark energy, e.g., by CMB anisotropy, the SDSS and 2dF
surveys, or gravitational
lensing, is crucial.  Additional measurements of SNeIa could
help shed light on the precise nature of the dark energy: there
are interesting possibilities beyond vacuum energy.  The dark
energy problem is not only of great importance for cosmology, but
for fundamental physics as well.  Whether it is vacuum energy or
quintessence, it is a puzzle for fundamental physics and likely
a clue about the unification of the forces and particles.

\item Present expansion rate $H_0$.  Direct measurements
of the expansion rate using standard candles, gravitational time
delay, SZ imaging and the CMB maps will pin down the elusive
Hubble constant once and for all.  It is the fundamental parameter
that sets the size -- in time and space -- of the observable Universe.
Its value is critical to testing the self consistency of Cold Dark Matter.

\item Dark matter bookkeeping.  Our best knowledge of the amount
of matter in the Universe is based upon clusters of galaxies.
Two new X-ray observatories -- NASA's AXAF and ESA's XMM -- will
be launched in 1999, and data they take will strengthen and refine
our understanding of dark matter based upon clusters of galaxies.
Further, a powerful new tomographic technique for studying
clusters when combined with x-ray measurements will sharpen
measurements of dark matter in clusters.  (The technique, Sunyaev --
Zel'dovich or SZ imaging, uses the fact that some fraction of
CMB photons that pass through a cluster have their energies
changed slightly.)  Until a decade
ago, almost all knowledge of the distribution of matter in the Universe
was based upon the distribution of light.  Gravitational lensing
by dark matter has begun to reveal the distribution of matter; this
technique, which requires CCD cameras with 100s of millions
of pixels and telescopes
with wide fields of view, will undoubtedly help us to better understand
the distribution of dark matter and test the Cold Dark Matter hypothesis
(Tyson, 1993).

\item Big-bang nucleosynthesis.  We should not forgot possible
insights that could come from more precisely probing the standard cosmology.
The Tytler -- Burles deuterium measurement and pegging of the density of
ordinary matter makes it possible to very precisely predict the
big-bang abundance of $^4$He, $24.6\%\pm 0.1\%$.  Current measurements
of the primeval $^4$He abundance are not nearly so precise,
$24\%\pm 1\%$.  Further measurements of the $^4$He abundance
have the potential to test this powerful probe of the hot big-bang
model and to strengthen the foundations of cosmology (or to shake them!).

\end{itemize}

\section{Looking Forward}

Because this is a debate, I have been purposefully provocative (as my
colleagues can testify, even more so than usual).
The scientist in me appreciates that we are still far from
``Cosmology Solved,'' and that the solution may be richer than or
even radically different from Inflation + Cold Dark Matter.
Big surprises could still be ahead.  Still, I think I can see
the top of the mountain emerging through the haze.

By any measure, Cosmology is entering a Golden Age, as prophesied
by David Schramm.  We have a well established foundation in
the hot big-bang model; we have bold and expansive theoretical ideas
born of the inner space / outer space connection, and now,
we are seeing the beginning of an avalanche
of high-quality observations that will test these ideas --
1998 was only the tip of the iceberg!

It may well be
-- and it is certainly my opinion -- that 1998
is remembered as the year that Inflation + Cold Dark Matter
became a part of the standard cosmology.  Or, it may be written
that 1998 was zenith for Inflation + Cold Dark Matter, and it was
downhill for it thereafter.  If the latter proves to be true,
armed with an enormous amount of information about the origin
and evolution of the Universe and with expectations for learning even
more, we will have to go back to the drawing board for new ideas.
And there is no doubt that those ideas will have to come from
the inner space / outer space interface.

If Inflation + Cold Dark Matter does pass the series of stringent
tests that will confront it in the next decade, there will
be questions to address and issues to work out.  Exactly how does
inflation work and fit into the scheme of the unification of the
forces and particles?  Does the quantum gravity era of cosmology,
which occurs before inflation, leave a detectable imprint on the
Universe?  What is the topology of the Universe?  Are there
additional spatial dimensions, and if so, how many and how
big?  Precisely how did the excess of matter
over antimatter develop?  What happened before inflation?  What
does Inflation + Cold Dark Matter teach us about the unification
of the forces and particles of Nature?
And then there is the amazing zoo of objects in the Universe to
understand.

\paragraph{Acknowledgments.}  I thank the other participants
in this Great Debate, for the many conversations, recent and over
the years, that have stimulated and helped shape my thinking about
the Universe.  My work at the boundary of elementary particle physics and
cosmology is supported by the Department of Energy (since 1978)
and by NASA (since 1983), for which I am very grateful.

\newpage

\begin{thebibliography} {cst}

\bibitem{} Bardeen, J., P. J. Steinhardt, and M. S. Turner, 1983,
Phys. Rev. D 28, 679.

\bibitem{} Bennett, C. et al, 1997, Physics Today, November, 32.

\bibitem{} Blumenthal, G., S. Faber, J. Primack, and M. Rees,
1984, Nature 311, 517.

\bibitem{} Burles, S. and D. Tytler, 1998a, Astrophys. J. 499, 699.

\bibitem{} Burles, S. and D. Tytler, 1998b, Astrophys. J., in press.

\bibitem{} Madore, B. et al, 1998, Nature, in press.

\bibitem{} Fixsen, D. J., et al, 1996, Astrophys. J. 473, 576.

\bibitem{} Fukuda, Y. et al (SuperKamiokande Collaboration) 1998,
Phys. Rev. Lett. 81, 1562.

\bibitem{} Geller, M., and J. Huchra, 1989, Science 246, 897.

\bibitem{} Guth, A., 1982, Phys. Rev. D 23, 347.

\bibitem{} Guth, A. and S.-Y. Pi, 1982, Phys. Rev. Lett. 49, 1110.

\bibitem{} Hawking, S.W. 1982, Phys. Lett. B 115, 295.

\bibitem{} Kolb, E. W., and M. S. Turner, 1990, {\it The Early Universe}
(Addison-Wesley, Redwood City).

\bibitem{} Krauss, L. \& M.S. Turner, 1995, Gen. Rel. Grav. 27, 1137.

\bibitem{} Linde, A., 1990, Inflation and Quantum Cosmology
(Academic Press, San Diego, CA).

\bibitem{} Ostriker, J. P., and P. J. Steinhardt, 1995, Nature 377, 600.

\bibitem{} Peebles, P. J. E., D. N. Schramm, E. L. Turner and R. G. Kron,
1991, Nature 352, 769.

\bibitem{} Perlmutter, S. et al, 1998, ApJ, in press.

\bibitem{} Riess, A. G. et al, 1998, Astron. J., in press (astro-ph/9805201).

\bibitem{} Sadoulet, B., 1999, Rev. Mod. Phys., in press.

\bibitem{} Schramm, D. and M. Turner, 1998, Rev. Mod. Phys. 70, 303.

\bibitem{} Silk, J. 1980, The Big Bang (W.H. Freeman and Co, San Francisco)

\bibitem{} Sloan Digital Sky Survey (SDSS), 1998, http://www.sdss.org

\bibitem{} Smoot, G. et al, 1992, ApJ, 396, L1.

\bibitem{} Starobinskii, A.A., 1982, Phys. Lett. B 117, 175.

\bibitem{} Turner, M.S. 1997, in Critical Dialogues in Cosmology,
ed. N. Turok (World Scientific, Singapore), p.~555.

\bibitem{} Two-degree Field (2dF), 1998, http://mso.anu.edu.au/~colless/2dF

\bibitem{} Tyson, J. A., 1993, Phys. Today 45, 24.

\bibitem{} White, S.D.M., C. Frenk, and M. Davis, 1983,
Astrophys. J. 274, L1.

\bibitem{} White, S.D.M. et al, 1993, Nature 366, 429.

\end{thebibliography}
\end{document}